\documentclass[final]{ustcstep}

\usepackage[dvips]{graphicx}
\usepackage[square]{natbib}

\def\gsim{\;\lower4pt\hbox{${\buildrel\displaystyle >\over\sim}$}\;}
\def\lsim{\;\lower4pt\hbox{${\buildrel\displaystyle <\over\sim}$}\;}
\def\grls{\;\lower4pt\hbox{${\buildrel\displaystyle >\over <}$}\;}

\newcommand\addr[2]{{\footnotesize \it $^{#1}$#2}\\}

\begin{document}

\title{Slow Magneto-acoustic Waves Observed above Quiet-Sun Region in a Dark Cavity}

\author{Jiajia Liu,$^{1,\dag}$ Zhenjun Zhou,$^{1,\dag}$ Yuming Wang,$^{1,*}$ Rui Liu,$^1$ Bin Wang,$^2$ Chijian Liao,$^1$ Chenglong Shen,$^1$\\[1pt]
Huinan Zheng,$^1$ Bin Miao,$^1$ Zhenpeng Su,$^1$ and S. Wang,$^1$ \\[1pt]
\addr{1}{CAS Key Laboratory of Geospace Environment,
Department of Geophysics and Planetary Sciences,University of Science \& Technology}
\addr{ }{ of China, Hefei, Anhui 230026, China}
\addr{2}{Beijing Institute of Tracking and Telecommunication Technology, Beijing, 100094, China}
\addr{\dag}{These authors contributed equally to this work}
\addr{*}{Correspondence and requests for materials should be addressed to Yuming Wang (ymwang@ustc.edu.cn)}}

\maketitle
\tableofcontents

\begin{abstract}
Waves play a crucial role in diagnosing the plasma properties of
various structures in the solar corona and coronal heating. Slow
magneto-acoustic (MA) waves are one of the important
magnetohydrodynamic waves. In past decades, numerous slow MA waves
were detected above the active regions and coronal holes, but rarely
found elsewhere. Here, we investigate a `tornado'-like structure consisting
of quasi-periodic streaks within a dark cavity at about 40--110 Mm
above the quiet-Sun region on 2011 September 25. Our analysis
reveals that these streaks are actually slow MA wave trains. The
properties of these wave trains, including the phase speed,
compression ratio, kinetic energy density, etc., are similar to
those of the reported slow MA waves, except that the period of these
waves is about 50 s, much shorter than the typical reported values
(3--5 minutes).
\end{abstract}

\section{Introduction}
Magnetohydrodynamic (MHD) waves are important phenomena in solar
corona and catch broad interests. They are believed to be the energy
carriers for coronal
heating \citep[e.g.,][]{Erdelyi2007, Jess2009, McIntosh2011},%, ***coronal heating***},
and also be a powerful tool to diagnose plasma properties and
physical processes in the solar atmosphere
\citep[e.g.,][]{Roberts2000, Banerjee2007, Marsh2009}. Slow
magneto-acoustic (MA) waves are one of the eigenmodes of MHD waves,
and were frequently observed above the active regions, e.g., in
coronal loops, and above coronal holes, e.g., in plumes
\citep[e.g.,][]{Ofman1997, Ofman1999, DeForest1998, Berghmans1999,
Berghmans2001, Schrijver1999, Nightingale1999, Moortel2002a,
Banerjee2007, Mariska2010}. The propagation speed of these waves is
usually around 100 km s$^{-1}$ with a period of about 3 min above
active regions or of about $10-15$ min above coronal holes. Due to
the different magnetic field configurations, they could be found
over a wide range of altitudes from 7 Mm to about several hundred
Megameters above coronal holes \citep[e.g.,][]{DeForest1998,
Ofman1997}, but usually below 20 Mm in coronal loops above active
regions \citep[e.g.,][]{Berghmans1999, Moortel2002a, Moortel2002b}.

In past decades, slow MA waves above quiet Sun regions were rarely
reported. It was probably due to the limitation of instruments.
Before the era of the Solar Dynamic Observatory (SDO) mission
\citep{Pesnell2012}, imaging instruments have provided either
full-disk observations at relatively low spatial resolution and low
cadence (e.g., SOHO/EIT, referring to \citealt{Delaboudiniere1995},
STEREO/EUVI, \citealt{Wuelser2004}), or high resolution and high
cadence but limited field of view (FOV) (e.g., TRACE,
\citealt{Handy1999}, Hinode/EIS, \citealt{Culhane1997}). The
instrument AIA \citep{Lemen2012} on board SDO provides so far the
best combination of resolution, cadence, FOV and EUV wavelength
coverage. In particular, its full-disk and round-the-clock coverage
allows one to launch a comprehensive search for a phenomenon of
interest. The only shortcoming of AIA data is the lack of Doppler
information. Although this shortcoming brings difficulty in the
identification of waves, it could be compensated by the analysis of
multi-wavelength observations. In this paper, we report an
observation of multiple slow MA wave trains in a dark cavity above a
quiet-Sun region, which may advance our understanding on the MA
waves in the solar atmosphere.

\section{Overview}
\begin{figure*}
\begin{center}
\includegraphics[width=\hsize]{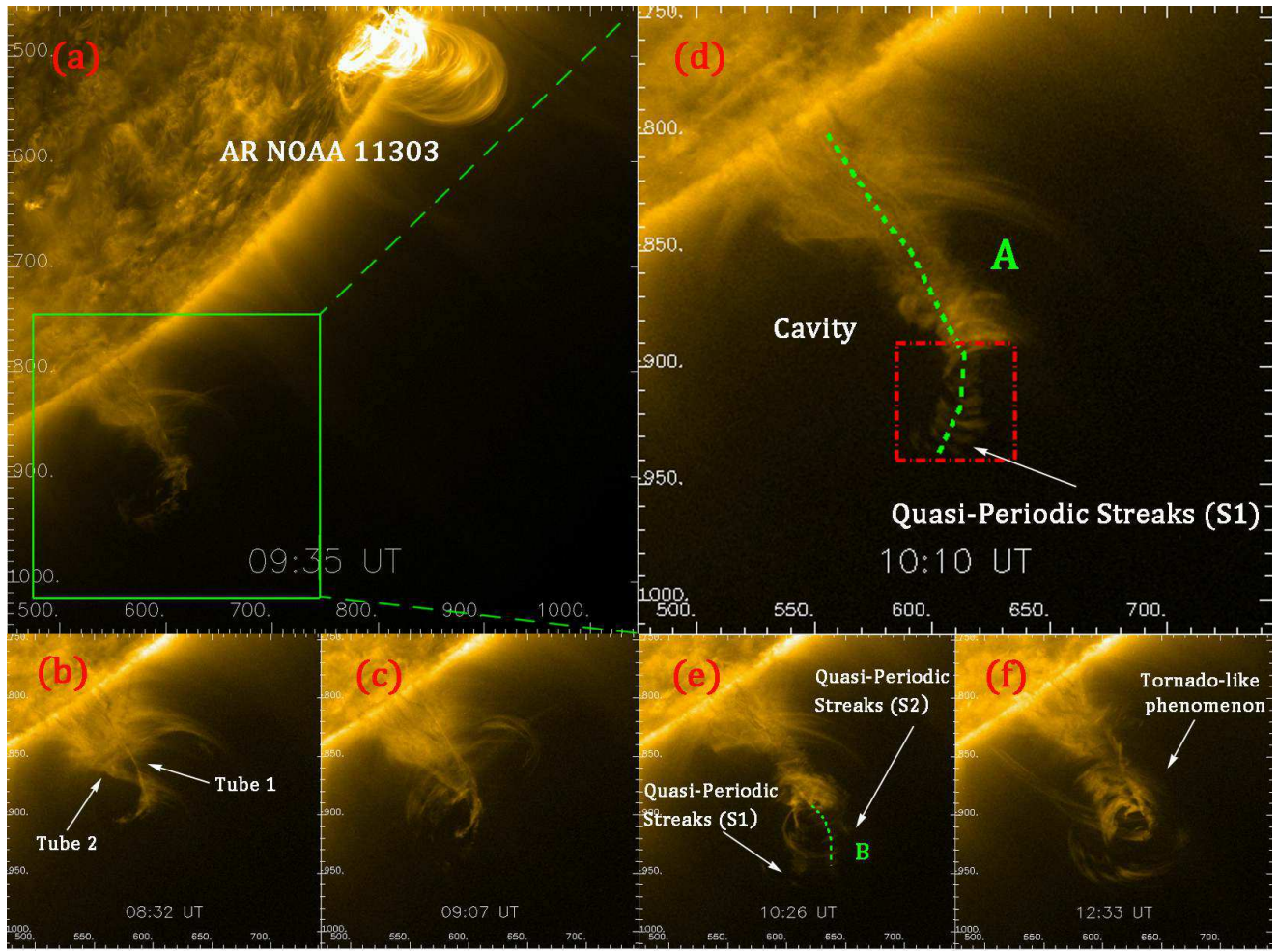}
\caption{Snapshots taken at AIA 171 {\AA} passband.
The FOV of image (a) is $600''\times600''$. Image (b)-(f) have the same FOV of
$270''\times270''$, which is just the region enclosed by the green box in image (a).} \label{ov}
\end{center}
\end{figure*}

The event of interest took place on 2011 September 25 above the
southwest limb within a dark cavity above the quiet Sun region (as
shown in Fig.~\ref{ov}(a)).
%The dark cavity with prominence material
%sitting on the bottom had been existed for a long time.
The overall
cavity structure stayed stable all the time, but the prominence-like
material under the cavity was perturbed at the beginning of
September 25 and finally settled down on the next day. The
process from 08:00 to 13:00 UT on September 25 can be seen in the
online movie M1. Among various phenomena in this process,
\citet{Li2012} reported a so-called `huge solar tornado'.

In this study, we focus on the nature of the repeated quasi-periodic
(QP) streaks starting from about 9:00 UT. The most clear group of QP
streaks (labeled as S1 in Fig.\ref{ov}(d)) appeared around 10:10 UT
in the 171 \AA\ channel, whose peak formation temperature is about
0.63 MK. Several 171 \AA\ images taken before, during and after this
interval are displayed in Figure~\ref{ov}. Before the appearance of
the QP streaks, there were apparently two plasma tubes (labeled as 1
and 2) standing right above the limb in the dark cavity
(Fig.~\ref{ov}(b)). Initially, the two tubes were slightly tangled
as if they were crossed each other in projection. Thereafter, those
two tubes were disturbed by an M1.5 flare in the nearby active
region NOAA 11303 (Fig.~\ref{ov}(a)) and ended up `apparently'
merging together after 9:20 UT.

Meanwhile, continuous upward flows along Tube 1 can be noticed to
originate from the tube bottom starting from around 08:20 UT. The
upward flows seemingly transformed into helical-like flows with
increasing height, and near the top of Tube 1 the helical-like flows
quickly expanded and developed into QP streaks, which intermittently
appeared with various intensity and broadness in the next 4 hours.
Figure~\ref{ov}(e) shows another group of streaks (labeled as S2).
These streaks propagated along the tube once they appeared.

\section{Helical magnetic structure, intermittent mass flows or wave trains?}
The nature of the QP streaks is of particular interest. Intuitively,
one may think that these QP streaks reflected the helical motion of
plasmas in a magnetic flux rope. If this conjecture is true, the
degree of the twist of the magnetic field in the flux rope could be
inferred from the number of streaks. In the 171 \AA\ image at 10:10
UT, for example, at least 10 streaks were displayed at the upper
portion of the tube, and it implies that the magnetic field lines
twisted more than 10 rounds or $20\pi$ in radian. Previous works
have suggested that a flux rope with force-free field tends to be
kink unstable when the twist exceeds about $3.5\pi$, and the growth
rate is on the order of minutes \citep{Hood1979, Torok2003,
Torok_etal_2004}. Some recent studies further suggested that,
for a flux rope with a larger aspect ratio or axial mass flow, the
threshold may increase up to $12\pi$ \citep{Baty_2001,
Srivastava_etal_2010, Zaqarashvili_etal_2010}. The QP streaks S1
studied here exceeded the threshold, but did not show any rapid
increase in size, brightness, or dramatic change in shape, which
means that there is no evident signature of the development of the
kink instability. Moreover, our data process cannot find any
signatures of the plasma motion along these streaks. Thus, these QP
streaks could not be the result of the plasma motion along twisted
helical magnetic field lines.

\begin{figure*}
\begin{center}
\includegraphics[width=0.8\hsize]{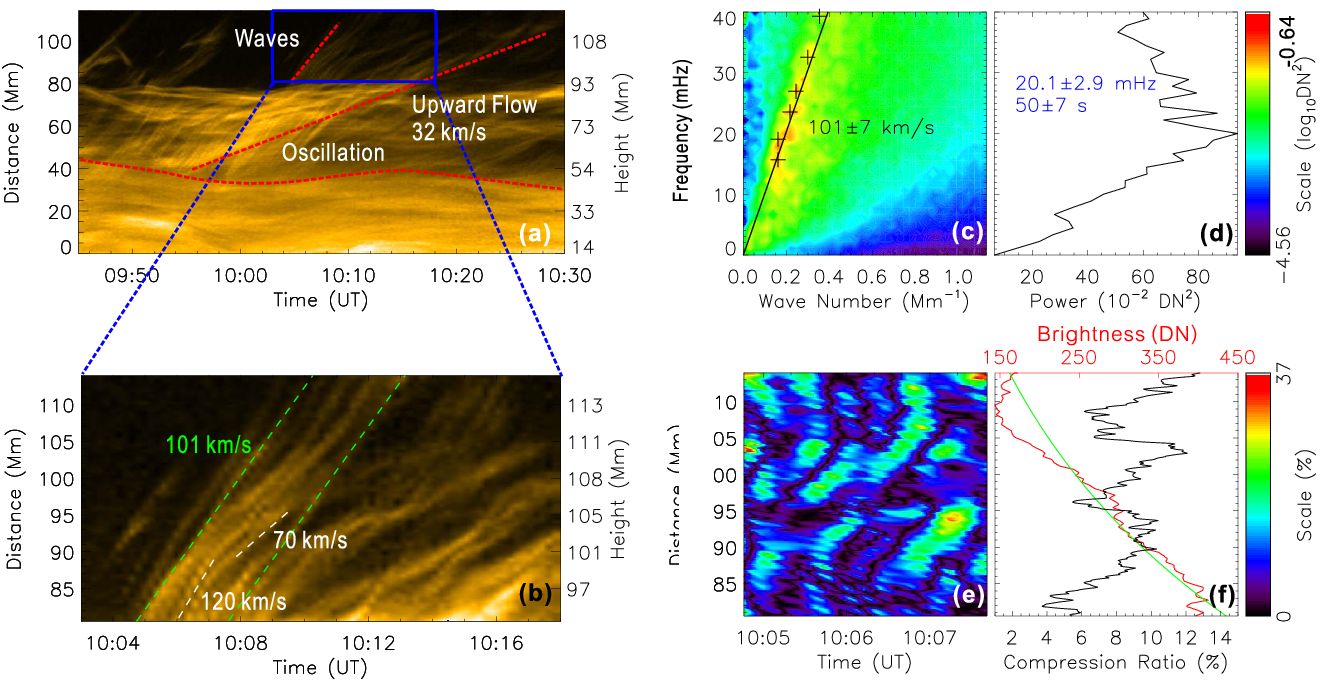}
\caption{(a) Space-time plot generated along Slice A from AIA 171
\AA\ images with a cadence of 12 s from 09:45 to 10:30 UT. (b) A
zoom-in plot of the region enclosed in the blue box in panel (a)
from 10:03 to 10:18 UT. (c) Fourier power distribution generated for
the 171 \AA\ data enclosed by the red box in Fig.\ref{ov}(d) from
10:03 to 10:18 UT. (d) Integrated power spectrum over wave number of
panel (c).  (e) Compression ratio in the region between the two
green lines in panel (b). (f) Average compression ratio over time of
panel (e). The red line shows the averaged brightness over time and
the green one is the fitting line.} \label{wave}
\end{center}
\end{figure*}

A wave train seems to be a more plausible interpretation for the QP
streaks. A 5-pixel wide slice, labeled `A' in Figure~\ref{ov}(d), is
placed along the axis of Tube 1 to analyze the plasma motion.
Figure~\ref{wave}(a) shows the space-time plot generated by
``seeing" through Slice A in AIA 171 \AA\ images. Its left vertical
axis gives the distance from the start point along the slice, and
the corresponding height from solar surface is marked in the right
vertical axis. It could be seen that the lower part of the tube
oscillated with a period of about 25 minutes, which was presumably
excited by the explosive activity in AR NOAA 11303 at around 09:30
UT (Fig.~\ref{ov}(a)). The upward flows mentioned before could be
identified as stripes with a positive slope in the plot
(Fig.~\ref{wave}(a)). Linear fitting to the stripes suggests that
the upward flows moved along the tube at a speed of about 32 km
s$^{-1}$.

The QP streaks S1 were located above the distance of 80 Mm after
10:04 UT in Figure~\ref{wave}(a). Figure \ref{wave}(b) shows a
zoom-in plot of the blue box region in Figure~\ref{wave}(a) from
10:03 to 10:18 UT. The temporal evolution of these streaks is shown
as alternating bright and dark stripes in the space-time plot.
Obviously, these stripes are steeper than those produced by the
upward flows. According to the space-time plot, the propagation
speed of these streaks varied from about 70 km s$^{-1}$ to 120 km
s$^{-1}$ with an average speed of about 100 km s$^{-1}$. It could
also be estimated that the spatial separation of the streaks is
about 5 Mm. Therefore, the period is about 50 s if the streaks were
caused by a wave train.

Following the method by \citet{Deforest2004} and \citet{Liu2011},
Fourier analysis is applied to S1. We generate a three dimensional
data cube using a series of running difference 171 \AA\ images in
the FOV enclosed by the red box in Figure~\ref{ov}(d) from 10:03 to
10:18 UT. The Fourier transformation converts the data cube from the
space-time domain $(x, y, t)$ to the wavenumber-frequency domain
$(k, \omega$) (as shown in Fig.~\ref{wave}(c)). A slanted stripe
with large powers (painted in yellow to red colors) can be clearly
seen in Figure~\ref{wave}(c). The phase speed of the wave can be
estimated from the slope of the stripe. For that purpose, we equally
cut the $k$-$\omega$ plot into six pieces along the frequency axis
above 10 mHz, and identify the point with the maximum power in each
piece (as marked by the plus signs in Fig.~\ref{wave}(c)). The phase
speed is just the average of the slopes of these points, which is
about $101$ km s$^{-1}$ with an uncertainty of about $7$ km
s$^{-1}$. The integrated power over wave number is presented by the
black line in Figure~\ref{wave}(d), which peaks at about $20\pm3$
mHz, corresponding to a period of about $50\pm7$ s. Both results are
consistent with the direct estimation from the space-time plot in
Figure~\ref{wave}(b).

Performing the same analysis to the QP streaks S2 shown in
Figure~\ref{ov}(e), we obtain a similar result which gives a phase
speed of about 110 km s $^{-1}$ and a period of about 45 s. These
waves were probably excited by the oscillation of the tubes or by
the upward flows from the tube bottom.

Recently, some argued that such QP streaks can also be explained by
heated QP mass flows \citep[e.g.,][]{DePontieu_McIntosh_2010,
McIntosh_etal_2010, DePontieu_etal_2011, Tian_etal_2011}. Those QP
mass flows may propagate into corona at a speed of order 100 km
s$^{-1}$. To distinguish a wave train from QP mass flows, we compare
the multi-wavelength observations from AIA, just like those done by
\citet{Kiddie_etal_2012}. If the observed QP streaks were produced
by mass flows, their propagation speed would not depend on the
temperature, i.e., there is no dispersion signature; otherwise,
the observed QP streaks were caused by a MA wave train.

\begin{figure*}
\centering
\includegraphics[width=0.8\hsize]{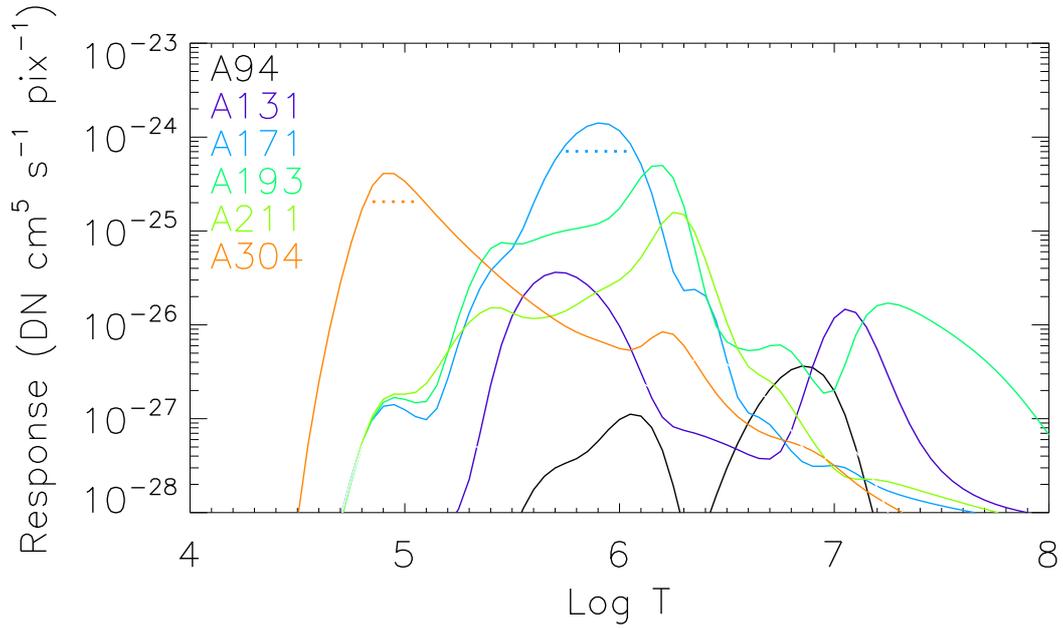}
\caption{Temperature response curves of AIA passbands. The
horizontal dashed lines give the full width at half maximum for 171
\AA\ and 304 \AA\ passbands.} \label{phase}
\end{figure*}

\begin{figure*}
\begin{center}
\includegraphics[width=0.8\hsize]{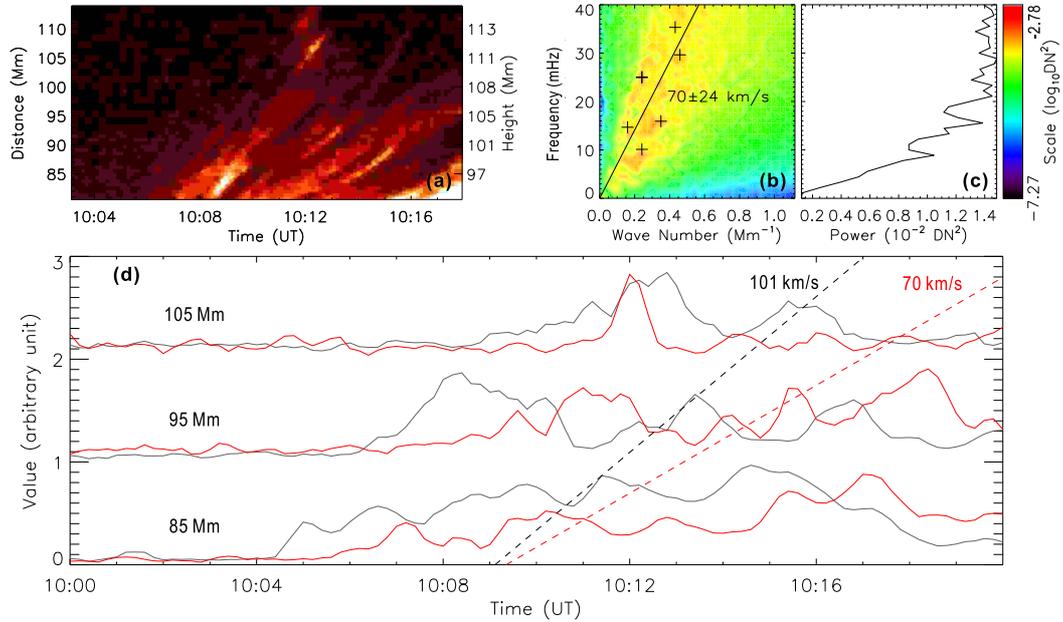}
\caption{(a)--(c) The same to
Fig.~\ref{wave}(b)--\ref{wave}(d), but for 304 \AA\ passband. (d)
Time sequences at distance 85, 95 and 105 Mm. The black line is for
the 171 \AA\ data, and the red line for 304 \AA. The dashed lines
guide the propagation of the same phase.} \label{wave304}
\end{center}
\end{figure*}

The temperature response curves of AIA passbands shown in
Figure~\ref{phase}, which is generated by the SSW procedure
`aia\_get\_response.pro' provided by AIA team \citep{Lemen2012},
suggest that, except 171 \AA\ and 304 \AA\ passbands, all the other
passbands obviously correspond to multiple temperatures. Thus, 304
\AA\ data is used to compare with 171 \AA\ data.
It is obvious that there is a similar
wave signature in 304 \AA\ data (Fig.~\ref{wave304}(a)--(b)), and even more importantly, the
phase speed is about $70\pm24$ km s$^{-1}$, which is much smaller
than the phase speed derived from 171 \AA\ data. However, there
is no clear peak frequency in the integrated power spectrum
(Fig.\ref{wave304}(c)); the power roughly stays at the same high
level from 20 to 40 mHz. This is probably due to the relatively low
signal-to-noise in 304 A data. Further, a comparison of the wave
phase between the 171 \AA\ and 304 \AA\ data is given in
Figure~\ref{wave304}(d), in which the time sequences at lower,
middle and higher positions are presented. The phase difference
between the two passbands varies with the position changing, which
means that the group of QP streaks S1 is likely a wave train rather
than mass flows. The dispersion of the MA wave suggests that
the plasma tube is likely a multi-thread structure, different
threads have different temperatures, and the propagation with
different phase speeds in different passbands is equivalent to the
propagation of waves in different threads.

\section{Evidence of slow magneto-acoustic waves}
The QP streaks as viewed in EUV 171 \AA\ reflect the fluctuation of
the emission intensity (i.e., brightness), and therefore the
fluctuation of the plasma density. It suggests that these waves are
compressional. In order to obtain the amplitude or compression ratio
of the wave train S1, the parallelogram defined by the two green
dashed lines in Figure~\ref{wave}(b), where the QP streaks are most
clear, is further investigated. The background (or unperturbed)
brightness in this region is a function of distance, which is
obtained by simply averaging the pixel values at each distance, as
this region covers several wavelengths of the wave. The red line in
Figure~\ref{wave}(f) shows the background brightness, $I$. The
amplitude of the fluctuation of the emission intensity, $dI$, is
therefore obtained by subtracting the background brightness from the
observed one. In optically thin corona, it is usually true that $I
\propto \rho^2$ and therefore the compression ratio, which is
defined as $|d\rho/\rho|$, could be approximately given by $|dI/2I|$
\citep{Aschwanden2004, Liu2011}.

Figure~\ref{wave}(e) presents the compression ratio in the region of
interest. Note that the parallelogram is rectified so that, along
the vertical axis, distance and time variations are coupled
together. Since the green dashed line is selected to match the wave
speed of 101 km s $^{-1}$, the vertical variation in
Figure~\ref{wave}(e) can be roughly interpreted as the variation of
the wave amplitude for certain phases. The black line in
Figure~\ref{wave}(f) gives the average compression ratio over time,
which is about 8.4\% on average with a standard deviation of about
2.0\%. No significant damping in amplitude is found over the
propagation distance of about 33 Mm. Since the amplitude of the
density fluctuation is relatively small compared to the background
density, it could be treated as a linear wave train. In linear
theory, of all propagating MHD waves, only MA waves can cause
density compression.

The region we studied is in a dark cavity, presumably void of
plasmas \citep{Low1996, Fuller2009}. It is hence expected that the
magnetic field inside the dark cavity is relatively stronger than
that outside, so that a pressure balance is maintained. Thus the
Alfv\'{e}n speed in the dark cavity should be approximately equal to
or even larger than the typical Alfv\'{e}n speed in quiet-Sun
region, which is about 220 km s $^{-1}$ \citep{McIntosh2011}. The MA
waves we investigated here propagated at much slower speeds than the
Alfv\'{e}n speed, and therefore these waves are slow MA waves.

Since these slow MA waves propagated along the tubes, which implies
that the wave vector is approximately parallel to the magnetic field
lines, the phase speed of the waves should be close to the sound
speed. The local sound speed is given by $c_s=\sqrt{\frac{2\gamma
kT}{m}}$, where $\gamma$ is the capacity ratio, $k$ is the Boltzmann
constant, $m$ is the proton mass and $T$ is the temperature. The 171
\AA\ passband corresponds to the temperature of about $0.56-1.1$ MK,
as indicated by the horizontal dashed line in Figure~\ref{phase}.
The sound speed at this temperature range is about $124-176$ km
s$^{-1}$. The 304 \AA\ passband corresponds to the temperature of
about $0.07-0.11$ MK, and the sound speed is about $44-56$ km
s$^{-1}$. These results are close to the phase speeds directly
derived from the imaging data, but not a perfect match, especially
for the 171 \AA\ data.  There could be various explanations
for the deviation. In a uniform magnetic field, if the wave
vector of S1 was not exactly parallel to the magnetic field, the
phase speed should be smaller than the local sound speed. Projection
effect also can result in the underestimation of the derived phase
speed from the imaging data. Besides, previous theoretical
analysis showed that, in a magnetic tube, slow MA waves propagate at
a so called tube speed, which is smaller than the local sound speed
\citep[e.g.,][]{Edwin_Roberts_1983}.

\section{Energy flux carried by the waves}
The kinetic energy density of a MA wave can be estimated by the
formula $\varepsilon=\frac{1}{2}\rho v_{ph}^2 (d\rho/\rho)^2$, where
$\rho$ is the plasma density and $v_{ph}$ is the phase speed of the
wave. The energy flux is given by $F=\varepsilon v_{ph}$. The
analysis of the 171 \AA\ data in the previous sections has revealed
that for wave S1, $v_{ph}$ is $101\pm7$ km s $^{-1}$ and
$d\rho/\rho$ is $8.4\%\pm2.0\%$. By further assuming the number
density to be the typical value of $10^8$ cm$^{-3}$ in dark cavities
\citep{Fuller2009}, we infer that the energy density carried by wave
S1 is $3.0\times10^{-5} \sim 1.2\times10^{-4}$ erg cm$^{-3}$, or the
energy flux is $2.8\times10^2 \sim 1.1\times10^3$ erg cm$^{-2}$
s$^{-1}$.

Since the emission intensity and therefore the density obviously
decreases with increasing distance (Figure~\ref{wave}(f)), the wave
energy actually dissipated when it propagated along the tube. We
applies an exponential equation $I(x)=ae^{c(x-b)}$ to fit the
observed brightness given by the red line in Figure~\ref{wave}(f).
The free parameters $a$, $b$ and $c$ are found to be 434, $-0.03$
and 80, respectively. The green fitting curve suggests that the
background brightness weakened by a factor of 63\% over a distance
of about 33 Mm or in about 5 min. Thus, it is inferred that the
average dissipation rate of the wave energy is $3.9\times10^{-8}
\sim 1.6\times10^{-7}$ erg cm$^{-3}$ s$^{-1}$.

\section{Summary}
\begin{figure*}
\begin{center}
\includegraphics[width=0.5\hsize]{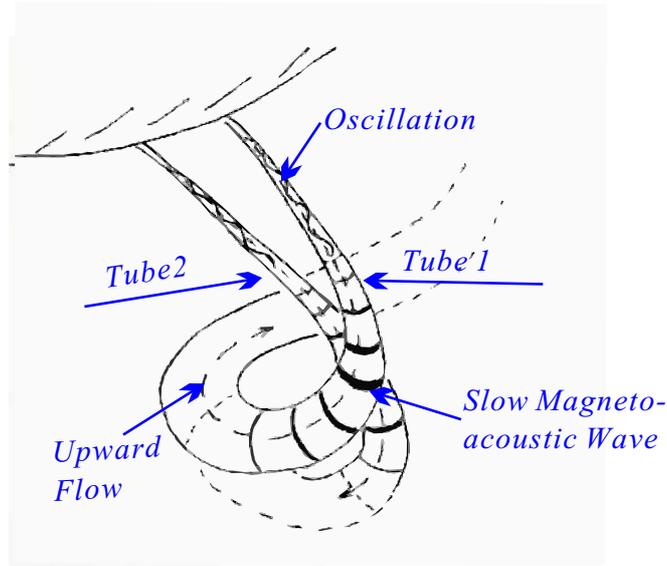}
\caption{An overview picture of the 2011 September 25 event.} \label{sum}
\end{center}
\end{figure*}

In this paper, we identified the QP streaks, appearing above the
quiet-Sun region on 2011 September 25, with the slow MA waves. The
overall picture of this `tornado'-like event is summarized in
Figure~\ref{sum}. There were two flux tubes in a dark cavity.
Oscillations and upward flows were initiated in the tubes around
08:00 UT by some perturbations. About one hour later, slow MA waves
were excited by the oscillations and/or the upward flows, and
manifested as QP streaks in the upper portion of the tubes. The
tornado-like phenomenon shown in Figure~\ref{ov}(f) was the results
of the bulk flows along the curved tubes.

The wave train S1 propagated along Tube1 around 10:10 UT is
particularly analyzed. Its propagation speed is about 101 km
s$^{-1}$ in the 171 \AA\ bassband or 70 km s$^{-1}$ in the 304 \AA\
passband, which is comparable with those of the slow MA waves found
above active regions, 122 km s $^{-1}$ \citep{Moortel2002a,
Moortel2002b} or coronal holes, $75-150$ km s $^{-1}$
\citep{DeForest1998}, while its period is about 50 s, which is much
shorter than those found above active regions, $\sim 3$ min, or
coronal holes, usually $10-15$ min. According to the fluctuation in
brightness, the density amplitude of the wave is about 8.4\%, also
comparable with that of the slow MA waves found elsewhere ($\sim4\%$
above active regions and $10\%-20\%$ above coronal holes,
respectively). The dissipation rate of the wave energy is estimated
from $3.9\times10^{-8}$ to $1.6\times10^{-7}$ erg cm$^{-3}$
s$^{-1}$, which is about one-order of magnitude smaller than that of
the radiative loss of the local corona \citep{Withbroe1977,
Aschwanden2006}.

Our study implies that (1) slow MA waves not only appear above
active regions or coronal holes, but also exist above quiet-Sun
regions, and the fact that they were rarely found above quiet-Sun
region before SDO mission is probably simply due to the instrument
limitation; and (2) the slow MA wave train in this study is not
sufficient for local coronal heating. It should be noted that
the slow MA waves were repeatedly observed during the event, and
sometimes, two or more wave trains appeared simultaneously. Whether
such slow MA waves ubiquitously exist in the corona and whether they
do a significant contribution to coronal heating is worth to be
investigated further.

\acknowledgments{We acknowledge the use of data from AIA instrument
on Solar Dynamics Observatory (SDO). This work is supported by
grants from the CAS (the Key Research Program KZZD-EW-01-4,
100-talent program, KZCX2-YW-QN511 and startup fund), 973 key
project (2011CB811403), NSFC (41131065, 40904046, 40874075, and
41121003), MOEC (20113402110001) and the fundamental research funds
for the central universities.}

\end{document}